# Multiple intercalation stages and universal $T_c$ enhancement through polar organic species in electron-doped 1T-SnSe$_2$

Hanlin Wu[†], Sheng Li[†], Wenhao Liu[†], Bing Lv*[†]

[†] Department of Physics, The University of Texas at Dallas, Richardson, TX 75080, USA

ABSTRACT

In this work, we report multiple intercalation stages and universal $T_c$ enhancement of superconductivity in 1T-SnSe$_2$ through Li and organic molecules cointercalation. We observe significantly increased lattice parameters up to 40 Å and dramatically enlarged interlayer distance up to ~11Å in Li and N,N-dimethylformamide (DMF) cointercalated SnSe$_2$. Well-separated cointercalation stages with different stacking patterns have been discovered by carefully controlled reaction time and concentration of solutions. These cointercalation stages are superconductors showing different superconducting signals. In addition, Li and various organic species such as Acetone, Dimethyl sulfoxide (DMSO) and Tetrahydrofuran (THF) have been cointercalated into SnSe$_2$ crystals, all of which show enhanced superconducting $T_c$ compared to solely Li intercalated SnSe$_2$. Our findings may provide more insight to effectively tune electronic structure of the lamellar structure through organic molecules co-regulation, and open a new strategy to engineer the physical properties of these layered materials by controlling their different intercalation stages.

1. INTRODUCTION

Chemical intercalation through van der Waals interlayer space is an effective method to tune structural, electronic, optical and magnetic properties of many layered materials and therefore open up exciting opportunities for diverse applications.[1–7] Small-sized inorganic ions, either as reductants involving alkali metal, alkaline-earth metal, transition metal or as oxidants such as halogen, could be used to control electron/hole doping and modify interlayer interactions while maintaining stability of layered structures in these materials.[5–10] The intercalation could be achieved through immersing materials into related chemical solutions, electrostatic gating, or more complex electrochemical effects.[1,2] These intercalations have led to observations of superconductivity in many types of material systems such as graphite[3,11], transition metal dichalcogenides (TMDs)[8,12–17], topological insulators[18], iron chalcogenides[19–21], and various quasi-2D and quasi-1D materials[22–29]. In alkali metal or transition metal intercalated TMDs, superconductivity is induced through suppression of CDW state or fine-tuned electron occupation of a relatively narrow $d$ band, while a dome-like $T_c$ dependence on the intercalation level could be achieved.[7,8]

Intriguingly, "charge neutral" water molecules and even larger organic molecules, often with small inorganic ions, could also be inserted into interlayer space. Compared to the intercalation, additional water or organic molecules often yield enhancement of $T_c$, as seen in $Na_x(H_2O)CoO_2$, $A_x(H_2O)_yTaS_2$, β-XNCl (X=Ti, Zr, Hf), and Fe chalcogenide systems[28–33]. The exact role of organic molecules on electronic properties of cointercalated phases could not be calculated as the positions of these molecules between the layers have not been known. Understanding mechanism for enhanced superconductivity, and some unusual doping-independent superconducting behaviors are yet to be resolved. In addition, one often cannot precisely control the number of organic species into the layers nor observe different intermediate intercalation stages with different layer stacking.

1T-SnSe$_2$ is a semiconductor with indirect band gap of 1.07 eV, which crystallizes in CdI$_2$-type structure (space group: *P-3m1*) with lattice parameter of ~6.15 Å along *c* axis and interlayer spacing of ~3.0 Å (Figure 1). Superconductivity has been reported previously through chemical intercalation[12,34–36], electrostatic gating[37,38], high pressures[39] and interface engineering[40,41]. Possible charge density wave (CDW) and signature of unconventional superconductivity have been suggested. Interestingly, there is no clear evidence between suppression of CDW and superconductivity in SnSe$_2$, which is different from other TMDs such as TiSe$_2$ and TaS$_2$ system..[8] In our previous work, Li intercalated SnSe$_2$ show non-bulk superconductivity with transition temperature of 4.5K ,while an enhanced T$_c$ up to 7.6K is obtained via organic molecule cointercalation[12]. On the other hand, indication of different intercalation stages had been noticed from X-ray diffraction patterns in some of the cointercalated samples, which trigger us carry out more comprehensive and detailed studies to figure out the exact nature and stacking orders of different intercalation stages, and how physical properties is related with these different intercalated stages.

Under this motivation, we carry out experiments which allow us to control and isolate different intercalation stages for Li and N,N-Dimethylformamide cointercalated samples Li$_x$(DMF)$_y$SnSe$_2$. Clear evidence of three intercalation stages of Li$_x$(DMF)$_y$SnSe$_2$ is observed from X-ray diffraction. The result from electrical transport and magnetic susceptibility measurements suggests similar T$_c$ but different superconducting signal at different cointercalation stages. A clear transport anomaly, akin to CDW state, emerge in transport experiment for intermediate cointercalated stages. We then expand this approach to study cointercalation with other different types of organic molecules. All the cointercalated samples show universally higher $T_c$ compared to the 4.5K of Li intercalated SnSe$_2$ sample.

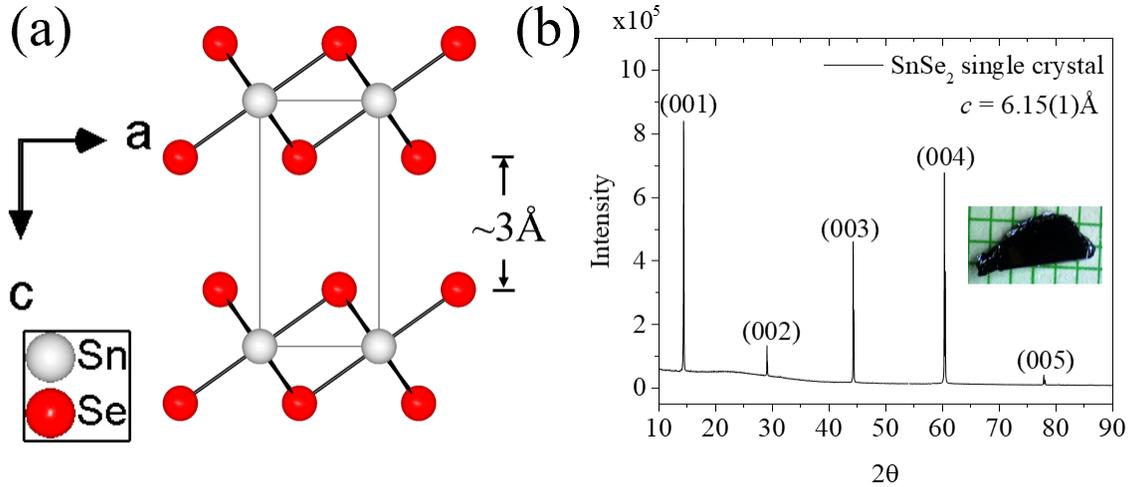

**Figure 1.** (a) Crystal structure of 1T-SnSe$_2$. (b) (*00l*) reflection peaks of XRD pattern for SnSe$_2$. Inset is the image of SnSe$_2$ crystal on a millimeter-scale sheet.

2. EXPERIMENTAL DETAILS

The high quality SnSe$_2$ single crystals are used for starting materials for cointercalation studies. They are grown through vertical Bridgeman method using pre-formed SnSe$_2$ powder synthesized from high purity tin (Alfa Aesar, 99.999%) and selenium (Alfa Aesar, 99.999%) in a doubled sealed quartz tube[12]. After the reaction, SnSe$_2$ crystals with typical size of 4×3×0.2 mm$^3$ are mechanically cleaved and used for cointercalation. All the X-ray diffraction, magnetic susceptibility, and resistivity measurements are performed on the as-grown or cointercalated single crystals. X-ray diffraction on the crystals show strong *c*-axis preferred orientation (Figure 1). The peaks at low angles, especially the strongest (001) peak, is used to compare and determine lattice expansion and cointercalation stages during the cointercalation process.

Since the samples after intercalation are air-sensitive, all the process are performed inside purified Ar-atmosphere glovebox with total $O_2$ and $H_2O$ levels < 0.1 ppm. N-butyllithium (n-BuLi) in hexane solution with 1.6 M concentration is purchased from Acros Organics and handled inside the glove box to avoid possible ignition or explosion. All organic solvents including Acetone, THF, DMSO, DMF, isopropanol (IPA) and propylene carbonate (PC), purchased from Alfa Aesar, are vacuum distilled and dried with molecular sieve before usage. The methods utilized for the cointercalation studies is described as following:

Lithium naphthalene in DMF solution (Li-Naph-DMF) with concentration of 0.05M, 0.1M, 0.15M and 0.2M were prepared by dissolving corresponding amounts of naphthalene and Li metal into DMF and heating at ~ 70 °C for 16 h. After obtaining Li-Naph-DMF solutions, the cointercalated $Li_x(DMF)_ySnSe_2$ were prepared by soaking $SnSe_2$ crystals in various concentration Li-Naph-DMF solution for different time.

When using other organic solvents, Lithium naphthalene could not dissolve completely in some organic solvents. Therefore, we develop another method for cointercalation. $Li_xSnSe_2$ samples were first prepared by soaking $SnSe_2$ single crystals in n-BuLi solution. Considering the thickness of mechanically exfoliated $SnSe_2$ single crystals is hard to control, which may affect reaction rate and experimental results, we therefore synthesize homogeneous cointercalated samples using sufficient reaction time (2 days) and high concentration of n-BuLi reactants (1.6 M solution). Then $Li_xM_ySnSe_2$ (M= Acetone, DMSO, THF, IPA and PC) crystals were obtained by soaking the resulting $Li_xSnSe_2$ in enough respective organic solutions for 2 days to reach the final intercalation stages.

All samples were dried out before measurements. X-ray diffraction patterns were collected using the Rigaku Smartlab, which is calibrated each time before running samples. Electrical resistivity ρ as function of temperature and magnetic field was measured by employing a standard 4-probe method, where four gold wires are attached on the surface of the sample parallelly using silver epoxy, down to 2 K in a Quantum Design Physical Property Measurement System. The magnetization is carried out using Quantum Design Magnetic Property Measurement System with applied magnetic field perpendicular to *c*-axis of the crystals.

3. RESULTS AND DISCUSSION

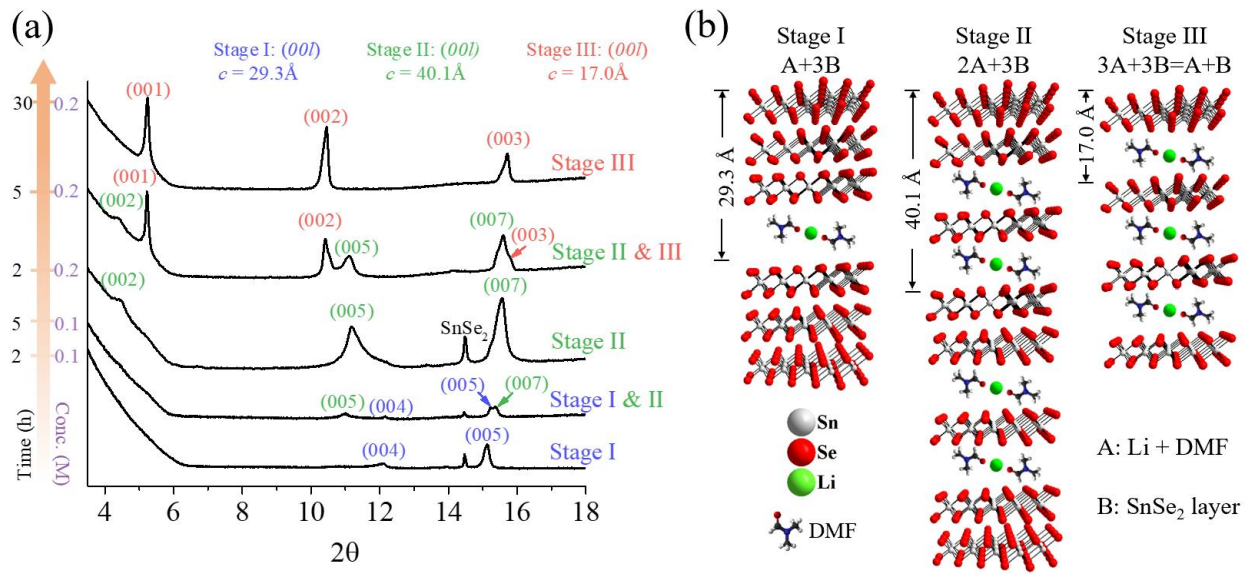

**Figure 2.** (a) (*00l*) reflection peaks of representative XRD patterns for $Li_x(DMF)_ySnSe_2$ at different intercalation stages. The detailed reaction time and concentration are listed on the left. (b) Systematic illustration of proposed crystal structure of Li and DMF cointercalated $SnSe_2$. The arrows and short breaks indicate unit cell of each intercalated stages.

Through precisely controlled experiments with different concentration of organic solutions and reaction time, we discover three different stages named as Stage I, Stage II and Stage III for $Li_x(DMF)_ySnSe_2$. Different intercalation stages are mainly separated by reaction time. For examples, Stage II is obtained using 0.2M solution with reaction time of 2 hours, and Stage III is synthesized using same solution but with reaction time of 30 h and beyond. Samples with reaction time in between 2h and 30 h normally have mixed Stage II and Stage III. The Stage I needs to be synthesized at lower concentration of 0.1M solution and with reaction time less than 2 hours. Noted that the exact reaction time to separate these intercalation stages is affected by thickness of crystals. Each sample is measured with X-ray diffraction firstly to confirm its own intercalation stage before we carry out the transport and magnetic studies discussed below.

**Table 1**. Detailed information from XRD pattern for Stage I, Stage II and Stage III.

| Stage | I | | | II | | | III | | |
|---|---|---|---|---|---|---|---|---|---|
| | (00$l$) | 2θ (°) | $d$ (Å) | (00$l$) | 2θ (°) | $d$ (Å) | (00$l$) | 2θ (°) | $d$ (Å) |
| Data from XRD pattern | (004) | 12.08 | 7.32 | (002) | 4.40 | 20.05 | (001) | 5.20 | 16.98 |
| | (005) | 15.10 | 5.86 | (005) | 11.06 | 8.00 | (002) | 10.42 | 8.48 |
| | | | | (007) | 15.46 | 5.73 | (003) | 15.64 | 5.66 |
| Estimated $c$ lattice | 29.3 Å | | | 40.1 Å | | | 17.0 Å | | |

Figure 2a shows the benchmark evidence of three intercalation stages from XRD patterns. Multiple peaks arise at the low angle range of SnSe$_2$ (001) peak. Through careful $d$ spacing and index calculations, as seen in Table 1, we are able to separate them into three sets of 00L peaks, corresponding to expanded lattice parameters of 29.3(2) Å for Stage I, 40.1(1) Å for Stage II, and 17.0(1) Å for Stage III.

The Li ion is rather small and could be easily inserted into the interlayer without noticeable lattice expansion[3], therefore, we could do some simple modelling of stacking order for these intercalation stages based on the size of SnSe$_2$ layer (~ 6.15 Å) and DMF organic molecules (10.85 Å, slightly larger compared to reported value[26]). The proposed structures with different stacking sequence is shown in Figure 1b. The unit cell of Stage I consists of three SnSe$_2$ layers and one layer of Li and DMF, so the basal spacing is 6.15 × 3 Å + 10.85 Å = 29.3 Å. For Stage II, three

SnSe$_2$ layers and two layers of Li and DMF form unit cell with basal spacing 6.15 × 3 Å + 10.85 × 2 Å = 40.15 Å. In Stage III, SnSe$_2$ layer alternatively stack with layer Li and DMF, which means completely reacted, and basal spacing is calculated as 6.15 Å + 10.85 Å =17 Å. This is reminiscent of different stacking stages observed in the graphite intercalation compounds but with different stacking order[3,6]. For instance, in K doped graphite, there are four different intercalation stages, named as Stage I, II, III and IV. Their structure consist of 4 layers of graphite/1 layer of potassium, 3 layers of graphite/1 layer of potassium, 2 layers of graphite/1 layer of potassium and 1 layers of graphite/1 layer of potassium, respectively.[3,6] It is noted that although we only discover three intercalation stages at present study in Li and DMF cointercalated SnSe$_2$ system, it is highly possible that more intercalation stages with different stacking orders could occur with changes of concentration, reaction time and even organic species.

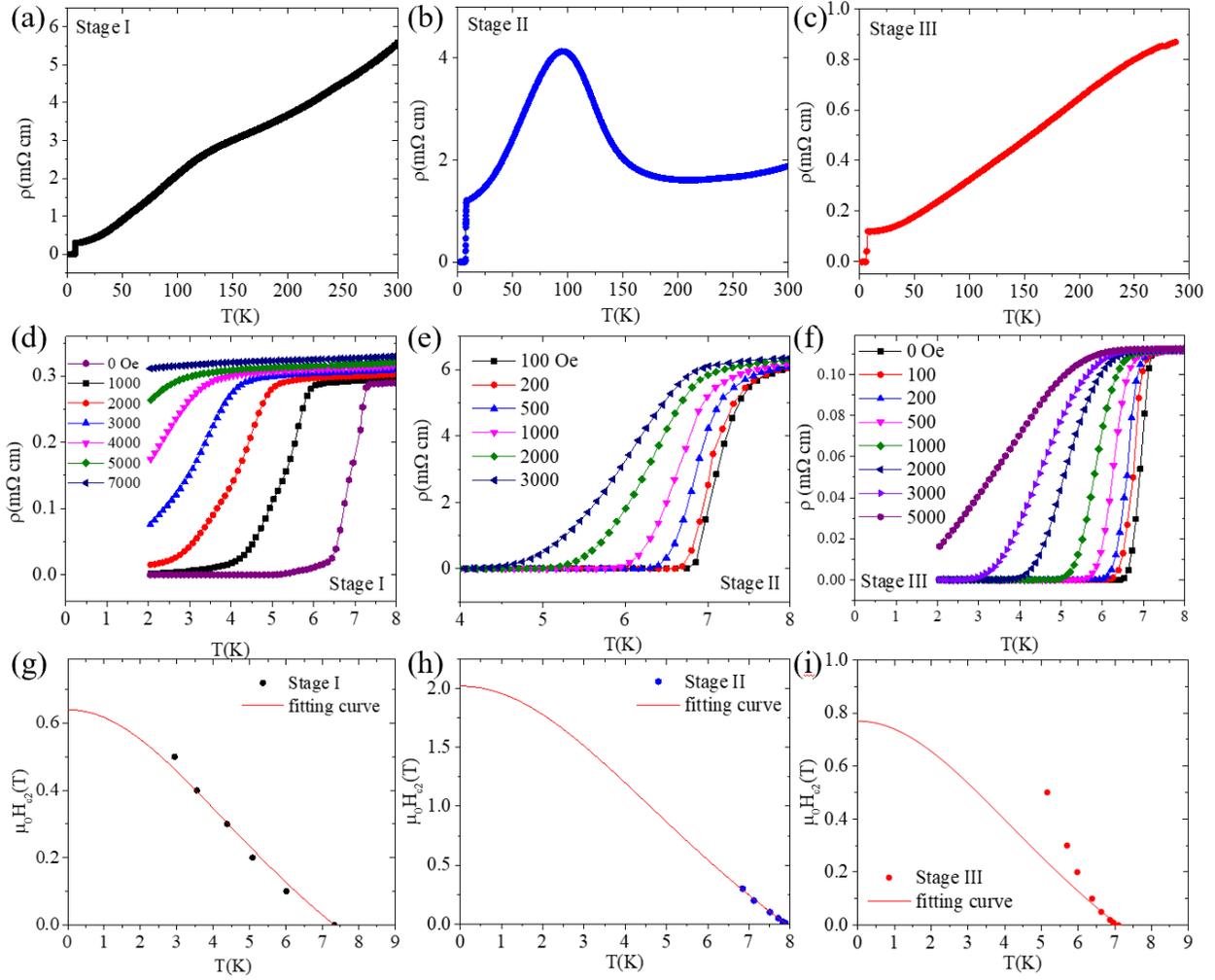

**Figure 3.** Temperature-dependent resistivity of $Li_x(DMF)_ySnSe_2$ for (a) Stage I, (b) Stage II and (c) Stage III; Field-dependent resistivity of $Li_x(DMF)_ySnSe_2$ (d) Stage I, (e) Stage II and (f) Stage III; Upper critical field determined from resistivity data by choosing cross point of two slopes for (g) Stage I, (h) Stage II and (i) Stage III. The red curve is fitting curve based on WHH model.

The success of isolation of three intercalation stages enable us to carry out temperature-dependent resistivity ρ(T) measurements of $Li_x(DMF)_ySnSe_2$ with Stage I, Stage II and Stage III to check their transport properties (Figure 3a-3c). In Stage I, a clear resistivity hump in ρ(T) curve is present at ~ 125K, which is likely associated with the CDW transition reported previously[34,42], and a superconducting resistivity drop at 7.2K is observed. In Stage II, superconducting transition temperature increases to 7.8K, while the CDW transition temperature is suppressed down to 100 K while the resistivity hump become much more pronounced. As both Stage I and Stage II show existence of CDW and superconductivity, in Stage III, the CDW transition appears to be completely suppressed and transition temperature is at 7.2K. Here the $T_c$ appears more or less constant regardless of appearance of the CDW state and additional amount of Li and DMF cointercalated into $SnSe_2$ layers at different intercalation stages. This behavior is different from other intercalated TMD materials, where a typical dome-shape $T_c$ with respect of charge carrier is observed, and superconductivity is induced by suppression (partially or completely) of CDW transition if there is any at normal states. In all three stages, the superconducting transitions are suppressed and get broadened upon the application of the magnetic field as expected, as shown in Figure 3d-3f. Interestingly, when we plot the upper critical field $H_{c2}$ as a function of transition temperature for Stage I, Stage II and Stage III (Figure 3g-3h), zero-temperature upper critical field, $\mu_0H_{c2}(0)$, of Stage I and Stage II are well fitted by the Werhmer-Helfand-Hohenberge (WHH) model, when only orbital effect is considered. And Stage III exhibit a very steep increase in the value of $H_{c2}$ with temperature and critical field as function of temperature derives significantly from WHH model. The $\mu_0H_{c2}(0)$ of Stage III is estimated to be 2.5 T following simple linear fitting, which is much higher than 0.8 T fitted by WHH model, suggesting likely multiband effect and unconventional superconductivity in Stage III when the system becomes more two dimensional as more Li and DMF molecules are cointercalated into layers of $SnSe_2$.

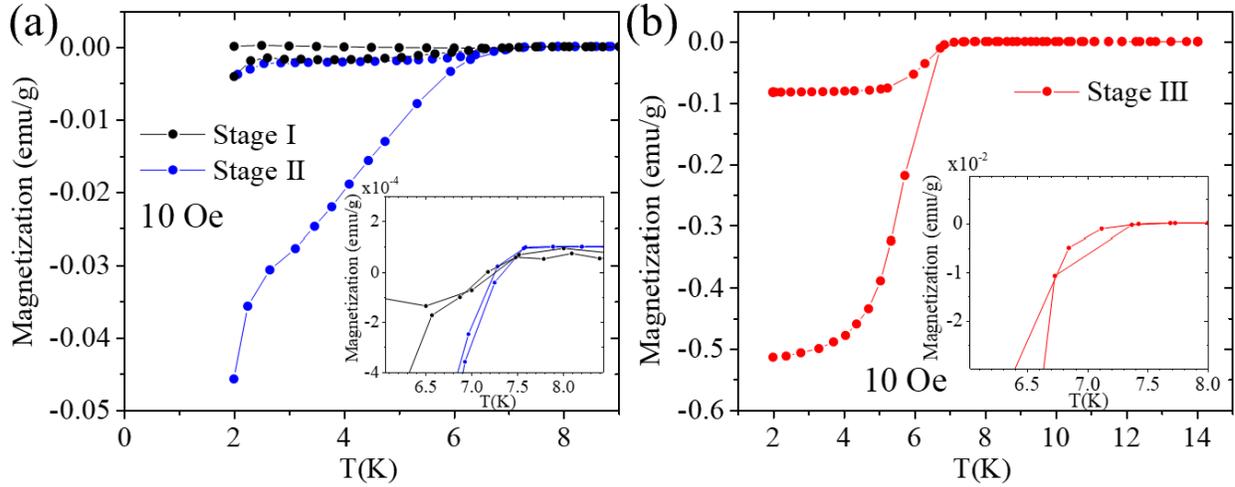

**Figure 4.** Magnetization data of Li$_x$(DMF)$_y$SnSe$_2$ with (a) Stage I and Stage II and (b) Stage III. Inset shows enlarged view around transition temperature for each stage.

Since resistivity behaviors under magnetic field are different for three intercalation stages, we then examine temperature dependent magnetic measurements of Li$_x$(DMF)$_y$SnSe$_2$ for Stage I, Stage II and Stage III. Without considering the demagnetization factor as the precise dimension determination is difficult due to the air sensitivity of the samples, both Stage I and Stage II show non-bulk superconductivity while bulk superconductivity exceed 100% volume fraction is observed in Stage III. It is noted though their demagnetization factor correction should be rather close to each other as all the samples are measured with H⊥c. As cointercalation progresses, superconducting volume fraction of Stage II is much larger (over ten times) than that of Stage I, while Stage III exhibits an even larger diamagnetic signal compared to other two intercalation stages (ten times than Stage II and ~100 times than Stage I). On the other hand, as shown in the insets of Figure 4, although the signal of Stage I is relatively weak, all three stages show a clear diamagnetic shift starting at around 7.6 K, showing nearly the same onset superconducting T$_c$. These results are consistent with data from resistivity measurement. Together with previous result that solely Li intercalation can only induce non-bulk superconductivity at 4.5K, our results suggest that polar organic solvent cointercalation such as DMF is effective for enhancing superconducting $T_c$ in $SnSe_2$ system.

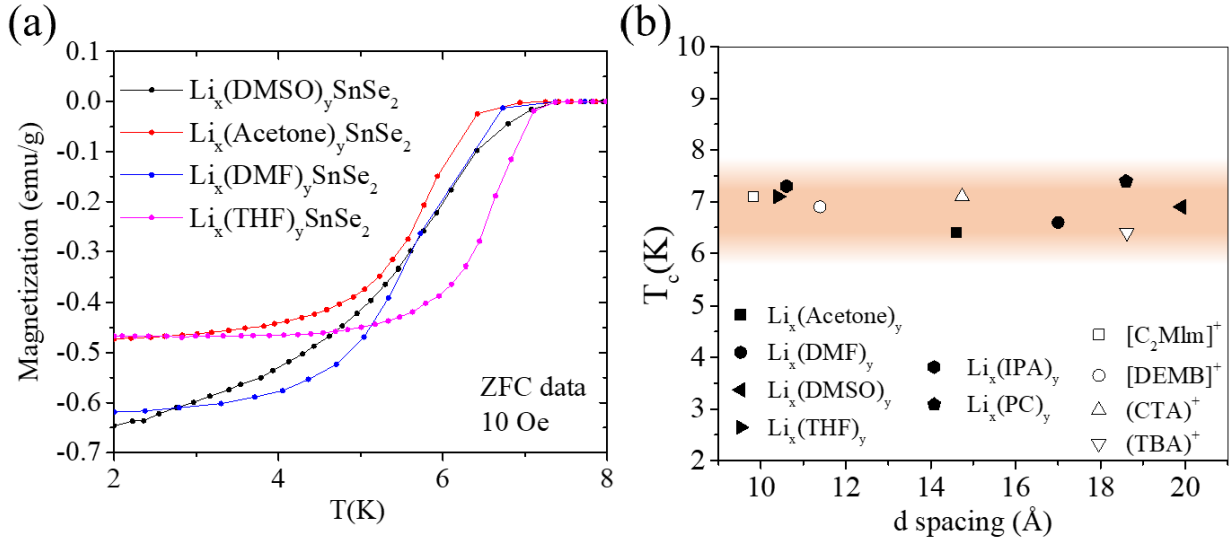

**Figure 5.** (a) Magnetization data of $Li_x(M)_ySnSe_2$, where M = DMSO, Acetone, DMF and THF. ZFC: Zero field cooling. (b) Superconducting $T_c$ values with respect to $d$ spacing distances in various cointercalated $SnSe_2$ superconductors. Black solid symbols: this work. Hollow symbols: reported data[35,36].

Since onset $T_c$ is independent of amount of cointercalated Li and DMF, it will be interesting to investigate the effect of other polar organic molecules on superconducting $T_c$ of cointercalated $SnSe_2$ system. Here by using the method mentioned in experimental section, $Li_x(M)_ySnSe_2$ (M= DMSO, Acetone and THF) samples are synthesized. Expecting onset $T_c$ won't change with the number of organic molecules, we focus on final cointercalation stage with enough soaking time (~2 days) and magnetic susceptibility characterizations. The results are shown in Figure 5a, where $Li_x(Acetone)_ySnSe_2$, $Li_x(THF)_ySnSe_2$ and $Li_x(DMSO)_ySnSe_2$ are all superconductors with high superconducting volume fraction exceed 100% volume fraction (also without demagnetization factor correction) with nearly the same onset $T_c$ at around 7 K. In addition, we have also tested other organic molecules such as isopropanol (IPA) and propylene carbonate (PC). Under similar soaking time of 2 days, these materials also become superconducting with similar $T_c$ but with volume fraction less than 20%, suggesting the cointercalation is not effective in these systems.

Since the size of organic molecules is significantly different from each other, we should expect different layer spacing (*d* spacing) from X-ray diffraction upon their cointercalation. To better understand the relationship between superconducting transition temperature and *d* spacing distances, we survey literature and plot a phase diagram together with our results as shown in Figure 5b. We choose the $T_c$s by cross point of two slopes from magnetic data to make the data consistent. The superconducting $T_c$s appear more or less independent from *d* spacing. A possible explanation for such relationship is that Li and organic molecules cointercalation has effectively increased basal spacing and two-dimensionality of $SnSe_2$, therefore subsequently manipulate charge and/or spin fluctuation, and enhance electron interaction/pairing for the increased $T_c$. The polar nature of organic molecules could contribute to the charge/electron transfer, but might only limited to those molecules close to the $SnSe_2$ layer. On the other hand, since the basal spacing between $SnSe_2$ layers are now large enough, interaction between these nearly two dimensional $SnSe_2$ layers now become weak, and therefore the $T_c$ will not be affected too much by further increasing *d* spacing distance.

## 4. CONCLUSION

In conclusion, chemical cointercalation through Li and different organic molecules has been carried out in $SnSe_2$ single crystals. For $Li_x(DMF)_ySnSe_2$ sample, we demonstrate three different interaction stages for the first time. Transport and magnetization measurements confirm superconductivity with $T_c$ at around 7.6 K in all three stages with different superconducting volumes. In addition, this unique method has been extended and universally applied to many other polar organic molecules. They all show increased $T_c$ comparing to solely Li intercalation, but the $T_c$s appear to be more or less the same with respect of organic molecules and the interlayer spacing caused by cointercalation. From our observation, the $T_c$ of cointercalated $SnSe_2$ is independent of doping concentration[12], cointercalation stages, organic molecules, and appearance or suppression of CDW transition, indicating this system might be an ideal platform to study the exact interplay of CDW, different cointercalation stages, and superconductivity, and further investigate the electron correlation effects for the related superconducting mechanism.

AUTHOR INFORMATION

**Corresponding Author**

*E-mail: blv@utdallas.edu.

ORCID

Hanlin Wu: 0000-0002-7920-3868

Wenhao Liu: 0000-0001-9757-1077

Notes

The authors declare no competing financial interest.

ACKNOWLEDGMENTS

This work at University of Texas at Dallas is supported by US Air Force Office of Scientific Research Grant No. FA9550-19-1-0037 and National Science Foundation (NSF)- DMREF-1921581.